\documentclass{article}

\usepackage{PRIMEarxiv}

\usepackage[utf8]{inputenc} 
\usepackage[T1]{fontenc}    
\usepackage{hyperref}       
\usepackage{url}            
\usepackage{booktabs}       
\usepackage{amsfonts}       
\usepackage{nicefrac}       
\usepackage{microtype}      
\usepackage{lipsum}
\usepackage{fancyhdr}       
\usepackage{graphicx}
\usepackage{tabularx}
\usepackage{wrapfig}

\title{Using Large Language Models for a standard assessment mapping for sustainable communities}

\author{
  Luc Jonveaux 
  \textit{(Mott MacDonald)}
}

\begin{document}
\maketitle

\begin{abstract}
This paper presents a new approach to urban sustainability assessment through the use of Large Language Models (LLMs) to streamline the use of the ISO 37101 framework to automate and standardise the assessment of urban initiatives against the six "sustainability purposes" and twelve "issues" outlined in the standard. The methodology includes the development of a custom prompt based on the standard definitions and its application to two different datasets: 527 projects from the Paris Participatory Budget and 398 activities from the PROBONO Horizon 2020 project. The results show the effectiveness of LLMs in quickly and consistently categorising different urban initiatives according to sustainability criteria. The approach is particularly promising when it comes to breaking down silos in urban planning by providing a holistic view of the impact of projects. The paper discusses the advantages of this method over traditional human-led assessments, including significant time savings and improved consistency. However, it also points out the importance of human expertise in interpreting results and ethical considerations. This study hopefully can contribute to the growing body of work on AI applications in urban planning and provides a novel method for operationalising standardised sustainability frameworks in different urban contexts.
\end{abstract}


\section{Introduction}

\subsection{Sustainable urban development challenges}

The 21st century presents urban planners and policy makers with unprecedented challenges. With over 55\% of the world's population living in urban areas and this figure expected to rise to 68\% by 2050, cities are at the forefront of global sustainability challenges. These range from climate change adaptation and mitigation to social equity and economic resilience. The complexity of urban systems, characterised by interdependencies between the built environment, social structures and ecological processes, requires sophisticated approaches to planning, delivering and monitoring, with a focus on the strategic planning phase to allow for strong strategic foundations \cite{freytag_living_2014}.  Traditional urban planning methods attempt to capture the complexity of sustainability in urban contexts, but the nature of urban governance, combined with the difficulty of quantifying and standardising sustainability metrics across different urban landscapes, does not allow for an easy, comprehensive and comparable assessment of urban sustainability initiatives. 

\subsection{Brief overview of ISO 37101 and its relevance}

In response to these challenges, the International Organisation for Standardisation (ISO) has developed the ISO 37100 standard series, entitled "Sustainable development in communities — Management system for sustainable development". This standard provides a structured framework for sustainable development in communities, offering a systematic approach to the planning, implementation and evaluation of urban sustainability initiatives through the implementation of management systems, from smaller scale systems like blocks and neighbourhoods \cite{codispoti_sustainable_2022} to bigger systems. At the conceptual centre of ISO 37101 is an approach defined by six sustainability 'purposes' and twelve families of city services, or 'issues'. This 6x12 grid provides a comprehensive lens through which places (being blocks, neighbourhood, cities, ..) projects, policies, strategies and initiatives can be viewed and evaluated. In particular, the strength of the standard suite of concepts is that it provides a common language and assessment framework for different contexts and scales, facilitating both local action and global comparability. However, its application is time-consuming, and requires the intervention of experts, preventing a more wide-spread application.

\subsection{Introduction to Large Language Models and their potential in standardized tasks}

Recent advancements in artificial intelligence (AI), have given rise to Large Language Models (LLMs). These models, trained on vast corpora of text data, demonstrate remarkable capabilities in understanding and generating human-like text across a wide range of domains. LLMs represent a paradigm shift in how we approach text-based tasks. Unlike traditional rule-based or statistical Natural Language Processing (NLP) methods, LLMs leverage deep learning architectures, typically based on transformer models, to capture complex linguistic patterns and semantic relationships. This allows them to perform tasks with a level of nuance and context-awareness previously unattainable in automated systems.

The potential of LLMs for repetitive tasks depending on clear, rigorous definitions is particularly interesting for us. Their ability to process complex, context-dependent information makes them well-suited for tasks that require both broad knowledge and specific domain understanding. In the context of assessing initiatives against a set group of definitions, LLMs have the potential to bridge the gap between qualitative descriptions of urban initiatives and standardized mapping, on frameworks such as the one defined by ISO 37101.

\subsection{Research objectives and significance}

This research aims to explore the interface between the structured framework of ISO 37101 and the capabilities of LLMs in the context of urban sustainability assessment. Specifically, we seek to:
\begin{itemize}
    \item Develop and test an approach to using LLMs to automatically categorise and assess urban sustainability initiatives according to the ISO 37101 framework.
    \item Explore the potential of this approach to break down silos in urban planning and enable more holistic, cross-sector sustainability strategies.
    \item Investigate the usability, scalability and adaptability of this method in different urban contexts and data sources.
\end{itemize}
The importance of this research lies in its potential to standardise and rationalise sustainability assessments in urban planning using LLMs, noting that this use case can be transferred to other assessments. The aim is to demonstrate the feasibility of a tool that can quickly process large amounts of unstructured urban project data and provide consistent, comparable assessments at a strategic level that are aligned with international standards. This could significantly improve the ability of cities to assess their sustainability initiatives, facilitate knowledge sharing between urban areas and ultimately accelerate progress towards achieving sustainable urban development goals. This study can also contribute to the growing body of work on the application of AI in standardised assessment frameworks. The insights gained from this study could influence future developments in both AI applications for urban sustainability and the development of international standards for sustainable urban development.

\section{Literature Review}

\subsection{Sustainable urban development and assessment frameworks}
The concept of sustainable urban development has evolved considerably since its beginnings in the late 20th century. Subsequently, frameworks such as the United Nations Sustainable Development Goals (SDGs) have set global targets, with SDG 11 specifically addressing sustainable cities and communities. Different initiatives such as CASBEE for Urban Development, LEED for Neighbourhoods, BREEAM Communities \cite{zhang_review_2019, pedro_integrating_2019, haapio_towards_2012} and DGNB Urban Districts \cite{pedro_systematic_2019} provide standardised criteria for assessing sustainable urban development at the neighbourhood level \cite{tam_green_2018, cegielska_green_2024, saiu_making_2022}, (often qualified as Neighborhood Sustainability Assessments - NSA \cite{elkamhawy_comparative_2024} and NSA Tools (NSATs) \cite{abastante_limits_2023}) which are helpful in evaluating initiatives to achieve certification \cite{feijao_comparative_2024}, but from a strategic perspective, these frameworks often focus primarily on specific (social or environmental) aspects and may overlook broader social and economic dimensions of sustainability

Strategic sustainability assessment of urban projects \cite{asaad_comparative_2024} remains a challenge due to inconsistent and incomplete data, difficulty in quantifying the qualitative aspects of sustainability lack of standardisation across different urban contexts \cite{riera_perez_multi-criteria_2013, sharifi_critical_2013, sharifi_neighborhood_2014, ameen_critical_2015} and the resource-intensive nature of comprehensive assessments, ultimately preventing services to citizen \cite{heaton_conceptual_2019} . These challenges emphasise the need for more efficient, consistent and adaptable assessment methods.

\subsection{ISO 37101: Structure, purpose, and applications}

The series containing ISO 37101 \cite{iso_iso_2016} , which was introduced in 2016, represents a more holistic approach to assessing urban sustainability, including mapping to a set of urban indicators \cite{iso_iso_2018}, specified for example in ISO37120 and operationalized in cities indicators datasets \cite{wccd_wccd_2014, midor_moving_2020, white_standardising_2021}. Its structure is based on six purposes (attractiveness, preservation and improvement of the environment, resilience, responsible resource use, social cohesion, and well-being) and twelve issues (from governance to biodiversity). provides a comprehensive matrix for assessing urban initiatives, as well as specific indicators \cite{zona-ortiz_novel_2023}. This approach was recently explored by the French public expertise agency for the ecological transition and regional planning (CEREMA) \cite{cerema_pourquoi_2020, cerema_comment_2020}. Thanks to the flexibility of the standard, it can be adapted to different scales, from individual projects to city-wide strategies. However, the relative complexity of the framework can make widespread adoption difficult, especially in resource-constrained urban environments

\subsection{Existing intersections of AI and sustainability assessment}

Urban planning and the AI domains are already intersecting \cite{wang_towards_2023}. More specifically, Studies such as \cite{peng_pathway_2023, sanchez_prospects_2023} have investigated the use of machine learning to predict indicators of urban sustainability. In the field of urban planning, LLMs have begun to find applications in policy analysis and public participation processes. These studies emphasise the potential of LLMs for processing and interpreting large amounts of unstructured text data, a common challenge in the context of urban planning. However, the specific application of LLMs to standardised sustainability frameworks such as ISO 37101 remains largely unexplored. This gap in the literature represents an opportunity for novel research that could significantly improve the efficiency and consistency of urban sustainability assessments.

The intersection of LLMs and the operationalisation of a standardised sustainability/urban development framework such as ISO 37101 offers potential solutions to several of the challenges identified in sustainability assessment. The ability of LLMs to process and interpret large amounts of textual data could address issues of data inconsistency and resource intensity of assessments, and their ability for nuanced language understanding could assist in quantifying qualitative aspects of sustainability.

\section{Methodology}

Our methodology is essentially based on the ISO 37101 framework, which provides a structured approach to assessing sustainability using a 6x12 matrix. The six sustainability purposes (attractiveness, preservation and improvement of the environment, resilience, responsible resource use, social cohesion, and well-being) form one axis of this matrix. The twelve sustainability issues (governance, education, innovation, health, culture, living together, economy, living and working environment, safety, infrastructure, mobility, and biodiversity) form the other axis. This methodology enables a comprehensive assessment of community initiatives by taking into account how each initiative contributes to the six purposes across the twelve issue areas.

We selected gpt-3.5-turbo as the primary LLM, with additional tests using other local models (llama2, mistral, ...). The choice of gpt-3.5-turbo was based on its proven ability to understand complex relationships and generate nuanced responses, as well as its ability to use “functions calling” to obtain JSON-structured data. This approach can later be ported to more recent models (in particular the latest OpenAI gpt-4o-mini model). Its large knowledge base, trained on various datasets, makes it well suited for interpreting different urban project descriptions from a general perspective. We also considered factors such as the accessibility of the API, the consistency of the results and the possibilities for fine-tuning. We tried locally-runned models as secondary models to allow to compare the results and qualitatively assess the generalisability of our approach across different LLMs, but rapidly discarded this because of the speed and usability across large datasets.

A critical component of this LLM-based approach the development of a custom prompt that summarises the essence of the ISO 37101 framework. This prompt was iteratively refined to guide the LLM in the use of the 6x12 framework, including a reworked, custom set of definitions for the 18 items. The prompt includes concise definitions of the six sustainability purposes and twelve issues, instructions for identifying relevant aspects of the input text, guidelines for assessing the project's contribution to each purpose-issue intersection, and a structured format for the output of the assessment results, qualified in the functions calling part of the request. Prompts have  been reworked through testing with a variety of urban project descriptions to ensure consistency and accuracy of LLM interpretations.

Finally, the processing code was written in Python, based on the Langchain toolbox, and FastAPI to provide a simple API framework. The texts were extracted and cleaned using standard text mining tools and processed in the following steps:

\begin{itemize}
    \item Pre-processing of the data: The project descriptions were cleaned and formatted for consistency.
    \item LLM analysis: Each project description is entered into the LLM along with our custom prompt, using OpenAI 'function calls' so to receive consistent, structured data corresponding to the 6x12 framework.
    \item Aggregation: Results are aggregated across projects to identify patterns and trends.
    \item Creation of visualisations and user-friendly reports in Excel format.
\end{itemize}

\section{Results}

\subsection{Application to two real world datasets}

\subsubsection{Selection of the datasets}

For the purpose of this experimentation, the experimental datasets comprises two different sources. These datasets were selected to represent a mix of citizen-led initiatives (Paris) and more structured, research-led projects (PROBONO), allowing us to test our methodology at different scales and contexts of  development.

\begin{enumerate}
    \item Paris Participatory Budget projects: subset of 527 projects from the City of Paris' open data portal, covering initiatives proposed and voted by citizens between 2014 and 2022.
    \item The PROBONO \cite{eu_integrator-centric_2022} project activities: review of a subset of 398 activities of the PROBONO Horizon 2020 project, which focuses on green building neighborhoods and living labs in six European cities, as well as the visions for the six living labs.
\end{enumerate}

\subsubsection{Qualitative analysis and control of LLM outputs}

Once the results had been produced in the form of visualisations and Excel-formatted reports, the LLM output was reviewed. The reviewed analysis was compared to the original analysis. Despite being limited to an original review sample of 26 initiatives with 177 data points, 80.1\% of data points were not changed by reviewers.  This revealed a high relevance of the categorisation of the initiatives, with possible creativity that the reviewers may not have had, and some ability to capture nuanced sustainability impacts that are not immediately obvious but still relevant. These figures of course will need to be updated with more reviews, as the project progresses.

\begin{figure}[htp!]
  \centering
  \includegraphics[width=.99\textwidth]{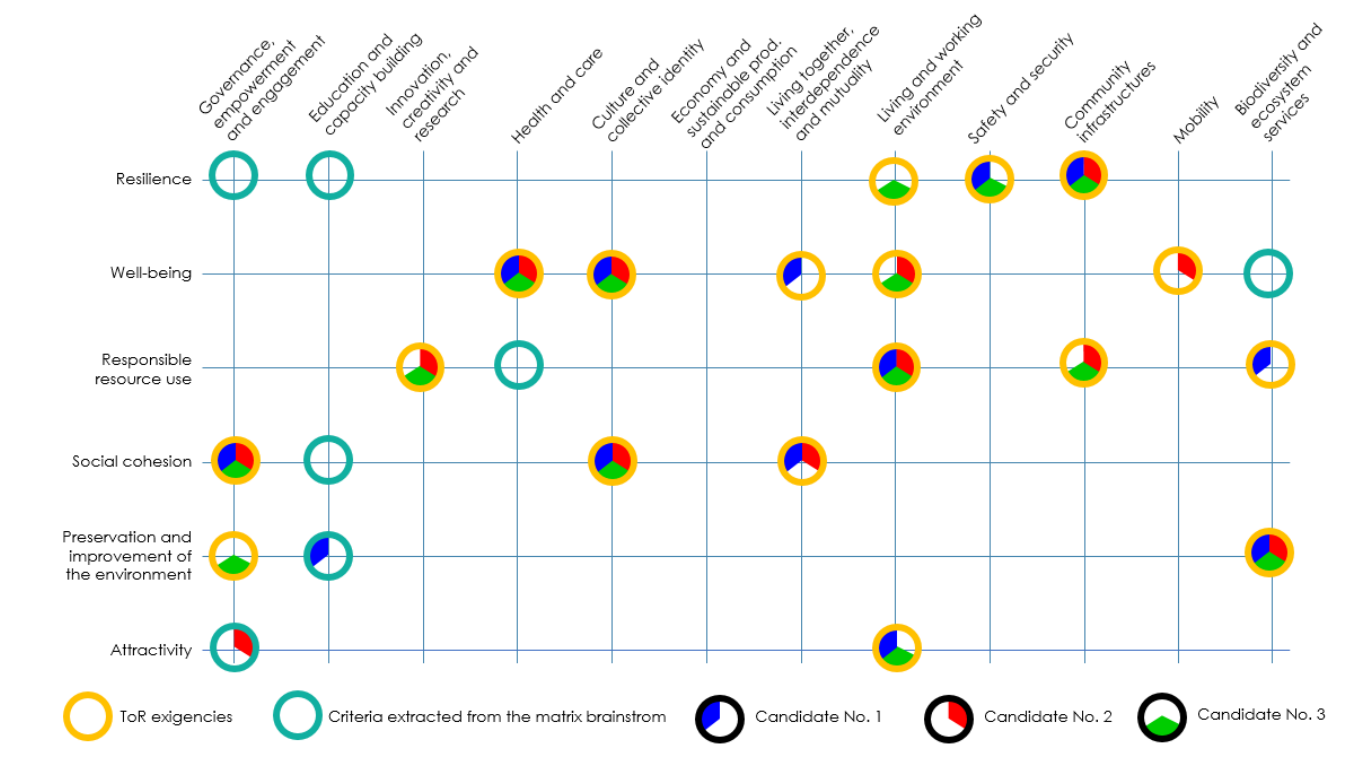}
  \caption{A review from an urban project in Grenoble, by Ecocités, showing alignment of tenderers to the needs of the city \cite{ecocite_iso_2020}. }
\end{figure}

\subsection{Findings from the analysis}

\subsubsection{Paris}

The Paris Participatory Budget is an participatory democratic initiative launched in 2014 that allows Parisians to propose and vote on projects to improve their city. With an annual budget of 100 million euros, it represents a significant commitment to citizen participation in urban development. We accessed the Paris Open Data portal \cite{paris_opendata_budget_2018} to obtain detailed information on the projects proposed and implemented between 2014 and 2022. Our data extraction script, written in Python using the Pandas library, collected project titles, descriptions, budget allocations and implementation status. The initial dataset contained 1759 initiatives, which we narrowed to 527 items, providing at least a description of the project longer than 100 characters.

These 527 projects cover 4271 positions on the 12x6 matrix, providing an average of 8.1 contributions (issue x purpose intersection), showing the interconnected impacts of such activities. This yields interesting observations, showing the priorities for Paris citizens. The main “Purpose” of the selection is “Social cohesion”, captured in 86\% of the projects, and Well-being, with 69.3\%. The main issues, in terms of number of initiatives, seem to be Living and working environment, with 68.8\% of the projects, and Culture and community identity, then Health and care in the community with 57.4\% and 57.1\% respectively. Issues like Production/consumption, Governance, Mobility and Innovation are below the 25\% threshold.

Moreover, based on this dataset, we developed a catalogue of "sustainable ideas" in which we have categorised exemplary projects under each purpose-issue overlap \cite{jonveaux_using_2024}. This can catalogue serve as a resource for urban planners and citizens and provides inspiration for future participatory budgeting proposals focused on sustainability goals.

\subsubsection{The PROBONO project}

PROBONO \cite{eu_integrator-centric_2022} is a Horizon 2020 project focusing on the development of Green Building Neighbourhoods (GBNs) and Living Labs in five European cities. We have analysed 398 activities from the PROBONO project, drawn from internal and external outputs, ranging from technical interventions to social innovation initiatives, as well as the visions and objectives of the project’s six Living Labs, which comprise 18 plans. Each activity and vision description was processed through our LLM-based analysis pipeline. The 398 activities and 18 plans of the project were mapped on the same framework, captured below. 

\begin{figure}[htp!]
  \centering
  \includegraphics[width=.99\textwidth]{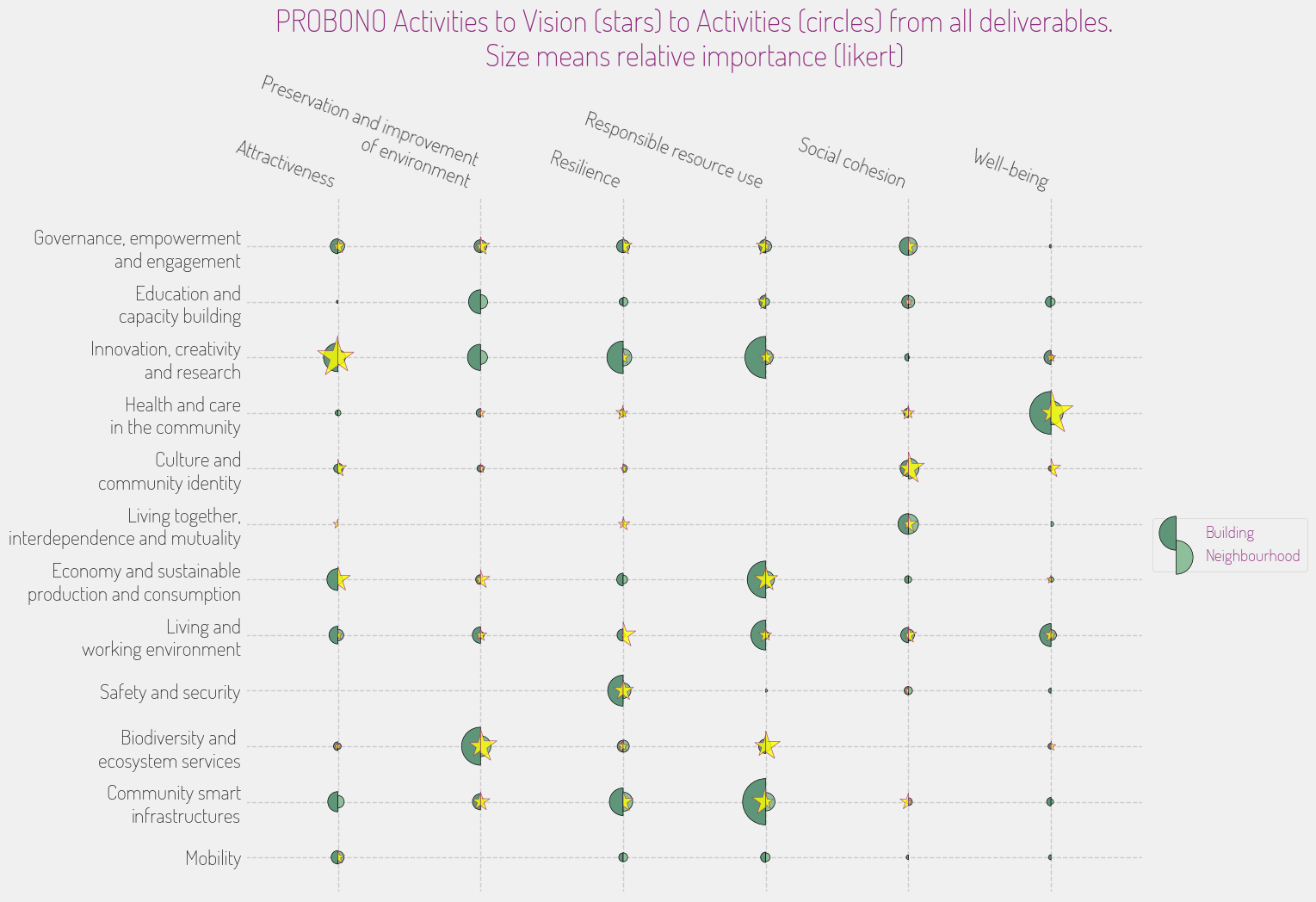}
  \caption{Review of the PROBONO \cite{eu_integrator-centric_2022} activities. }
\end{figure}

Stars represent the plans of the living labs, while the circles capture the essence of the activities.

This review helped highlight several aspects: 

\begin{itemize}
    \item Holistic approach: PROBONO activities showed a slightly less balanced distribution across the sustainability purposes compared to the Paris projects. This could be due to a smaller number of projects.
    \item Urban services: Activities scored particularly high in the "Innovation" and “Community Smart Infrastructure” issues,with 53.6\% and 47.1\% of projects respectively. Lower scores were observed on Mobility, Living together, Culture, and Education (10.5\%, 12.2\%, 11.5\% and 13.7\% respectively).
    \item Sustainability purposes observed: 74\% of projects aligned with “Responsible resource use” and a secondary purpose was Resilience at 59.1\%.
    \item Alignment of visions/activities: there is a certain alignment of visions and action, still, gaps between individuals plans and planned activities are helpful in identifying possible strategic adjustments to be made.
\end{itemize}

The results of the classification were shared with PROBONO demonstrators to give them a standardised view of how their activities aligned with the broader sustainability goals. This facilitated discussion on potential gaps and opportunities to improve the overall impact of the project.

\section{Discussion}

\subsection{Effectiveness of LLMs in standardized sustainability assessment tasks}
Results show that LLMs, when properly guided by a well-structured prompt based on the ISO 37101 framework, can effectively perform standardised sustainability assessments. This accuracy has yet to be quantified on this dataset, but it has been shown qualitatively that the ability of LLMs to process and interpret complex, contextualised information is particularly valuable in the field of urban sustainability, where projects often have multiple impacts. This capability addresses a key challenge of traditional assessment methods, which are often unable to capture the nuanced and interconnected nature of sustainability initiatives. However, it is important to recognise that the effectiveness of LLMs is highly dependent on the choice of model, the quality of the input prompt and the training data. The iterative refinement process we employed was critical to achieving high performance and emphasises the need for expertise in developing and fine-tuning such systems. 

\subsection{Comparison with traditional human-driven assessment methods}
The advantages of an LLM-based assistant versus fully human reviews are twofold. The first is the significant time saving compared to traditional manual assessments (a few seconds compared to around 30 minutes per project for an overview of the project). This efficiency gain opens up opportunities for more comprehensive and frequent sustainability assessments, allowing communities to assess a much wider range of initiatives and update their assessments more regularly. The second aspect is the consistency of automated assessments, which offsets another limitation of manual methods, namely the potential variability between assessors. This approach provides a standardised interpretation that can be particularly useful for comparing projects in different contexts or time periods. LLM have also enabled the use of virtual citizens \cite{zhou_large_2024}, with obvious limitations. Nevertheless, the use of AI in urban planning decision-making processes raises important ethical considerations \cite{urban_little_2021}. It is essential to ensure that LLMs complement, rather than replace, human expertise. Furthermore, while LLMs are excellent at recognising explicit relationships, they occasionally overlook subtle, context-specific factors that can be easily identified by human experts. There is therefore a risk that important qualitative factors or local knowledge that cannot be captured in standardised frameworks will be marginalised.

\subsection{Potential for breaking silos in urban development projects}
The visual representation of the project's impact could serve as a valuable communication tool, helping stakeholders to agree on a common understanding of the project's contributions to places and breaking down silos between areas of activity. The intention of this mapping is to highlight the synergies and trade-offs that may not be visible when projects are assessed in isolation by different stakeholders. This holistic view encourages integrated approaches to urban development and encourages collaboration between different departments and disciplines. For example, the link we found between social cohesion projects and biodiversity issues could encourage collaboration between stakeholders from social services and environmental planning departments. Furthermore, for large-scale projects such as PROBONO, the approach provides a powerful tool to ensure alignment between different activities and overarching sustainability goals. The ability to consistently categorise different activities according to a standardised framework can help project managers to identify gaps, redundancies and opportunities for synergies.

\subsection{Applicability to strategic reviews}
It is possible to map the vision of a place with its corresponding objectives through the lens of our LLM-based assessment framework, which provides urban planners and policy makers with a tool to identify gaps and assess strategic direction. It is possible to create a comprehensive visualisation that shows how a place's vision aligns with its specific programme of activities and planned interventions. This process enables the identification of areas where the vision is not sufficiently reflected by the current objectives, or conversely, where the objectives do not directly contribute to the overarching vision. Such mapping can reveal unexpected synergies or potential conflicts between different aspects of urban development strategies. It also provides a structured approach to assessing the holistic nature of a city's sustainability efforts and ensures that all dimensions of urban sustainability are adequately addressed. This strategic overview not only highlights areas that need to be improved or recalibrated, but also serves as a communication tool for stakeholders, enabling a shared understanding of the city's path towards its sustainable development goals. By using the map to compare the elements of the vision with the targeted actions on the map, responsible stakeholders can make more informed decisions, reallocate resources more effectively and ultimately develop more coherent and impactful sustainable development strategies.

\subsection{Applicability to solutions identification}

The systematic mapping of needs that may result from a gap analysis opens up opportunities to identify and utilise relevant activities from other contexts. Once the specific needs of a community are clearly articulated and mapped into the grid of sustainability goals and issues, this structured information can be used to interrogate a wider database of analysed and stored initiatives, for example based on the Paris Participatory budget catalogue of initiatives mentioned earlier.  Using the standard categorisation, users can identify analogous projects or measures that have been successfully implemented in other places. A simple cosine-based similarity score (based on a likert score, applied to the 144-long vector categorizing the 12 issues, 6 purposes, and 2 scales) has been successfully applied to sort solutions, based on its colinearity with the needs profile. 

This exchange of ideas and solutions can be powerful as it is based on the overview understanding of city services and sustainability impacts rather than superficial similarities. For example, a community's need for greater social cohesion around environmental preservation may lead to the discovery of innovative green space initiatives that have fostered community engagement elsewhere, but also identify already existing projects, for which the 'other' impacts were not previously considered. This approach can accelerate the learning process for communities stakeholder, and the adaptation and scaling of proven solutions in different urban contexts. It also encourages a collaborative and globally focssed approach to urban problem solving, where communities can learn from and build on the experiences of others. 

\section{Future Directions}

\subsection{Further work with urban planners, policymakers, and sustainability professionals}

The implications of our research go far beyond  theoretical discourse and offer considerable practical benefits for various actors in urban development. In particular, urban planners can benefit significantly from the rapid and consistent assessments that our approach enables. This methodology enables a more comprehensive assessment of proposed projects, facilitates data-driven decisions and promotes more holistic urban development strategies. This standardised, efficient sustainability impact assessment methodology allows planners to consider a wider range of factors in their designs, which can lead to a more resilient and sustainable urban environment.

Policy makers can also use this method to design more responsive and adaptive policies. The ability to quickly analyse large datasets of urban initiatives can provides a deeper insight into the collective impact of different programmes and projects. This capability can support an efficient alignment of local actions with the broader sustainability goals at the neighbourhood or city level, and international commitments, so to ensure that urban policy is both locally relevant and globally responsible.

For sustainability professionals, this provides a standardised framework and automated assessment tool that can be used as a common language for discussing and comparing different sustainability initiatives. This common lexicon has the potential to improve collaboration and knowledge sharing across different urban contexts, breaking down silos and promoting a more integrated approach to urban sustainability challenges.

The educational aspect of our research also opens up new avenues for specialists in the field of social engagement. This offers innovative tools for greater public participation in urban planning processes,  translating complex sustainability concepts into accessible, interactive formats. This approach could democratise urban planning by making it more inclusive and representative of the diverse needs and aspirations of the community, and can be leveraged in workshops similar to the Climate Fresk, promoting citizen and communities engagement.

Last but not least, decision makers involved in urban development projects can benefit from the detailed impacts categorisation  that the methodology provides. This detailed understanding can help align different activities with overarching sustainability goals, identify potential synergies between projects and identify unintended negative consequences, providing a more nuanced view of project impacts for more strategic and effective project management in the context of communities sustainability.

\subsection{Scaling the approach to larger urban systems and networks of cities}

Future research on this LLM-based assessment methodology should focus on adapting it for the assessment of city-wide or regional sustainability strategies, developing frameworks for cross-city comparisons and creating a global database  or 'catalogue' for urban sustainability initiatives. Still, to achieve scale, the relevance, accuracy and inclusiveness of sustainability assessments need to be improved and critical non-technical challenges addressed. Cross-cultural applicability needs to be explored, which may require localised models or prompts. The increasing role of AI in urban planning requires robust ethical guidelines and governance frameworks to ensure responsible and equitable use. As part of this endeavour, participatory AI methods that incorporate local knowledge and citizen input into our assessments should take centre stage. This could include platforms for citizen feedback on LLM-generated assessments, collaborative filtering systems that combine machine learning with crowd-sourced assessments, and the use of LLMs to analyse qualitative feedback from citizen participation.

\subsection{Refinement of LLM prompts and fine-tuning for specific contexts}

The promising results of our general LLM-based approach highlights the  potential for improvement through context-specific fine-tuning. Future research could focus on developing region-specific prompts that incorporate local sustainability priorities and terminology. This is essential for the accurate interpretation of initiatives in different urban contexts, e.g. the development of multilingual models could facilitate valuable cross-border knowledge exchange on urban sustainability practises. Moreover, fine-tuning the LLMs to domain-specific corpora of urban planning knowledge and sustainability reports will improve the model's understanding of technical concepts and lead to more nuanced assessments. The system could be improved with "few-shot" and "zero-shot" learning techniques to improve the model's stability and consistency. It would also make sense to explore smaller, specialised models that focus on urban sustainability, or using compact models could be trained or fine-tuned on curated urban sustainability datasets, which could provide better performance and accessibility for planning departments with limited resources.

\subsubsection{Development of a standardized toolkit for cities to self-assess their sustainability initiatives}

To facilitate wider application of our approach, future work should focus on creating a comprehensive, user-friendly toolkit for cities. This could include a web-based interface for submitting project descriptions and receiving automated  assessments, using customisable reports templates that align with common sustainability reporting frameworks. It could be accompanied with training materials and best practise guides for interpreting and applying the assessments produced by LLM, as well as tools to track progress over time and set data-driven sustainability targets. Such a toolkit could democratise access to sophisticated sustainability assessment techniques and enable cities of all sizes to benefit from AI-powered, human-led insights. The synergies with methods such as the “\textit{Fresque du Climat}” are obvious and could lead to a “\textit{Placemaking Fresk}” methodology.

\section{Conclusion}

Our research has shown the considerable potential of integrating Large Language Models (LLMs) with the ISO 37101 framework to change and accelerate the development of the strategic assessment of urban sustainability. The results of this study are both promising and multi-faceted, offering a new paradigm for evaluating and understanding urban development initiatives through the lens of sustainability.

The high accuracy and consistency of our LLM-based approach, validated by human-reviewed assessments, highlights its effectiveness in categorising urban projects across a spectrum of sustainability purposes and issues. This level of performance suggests that AI-powered assessments can be a reliable alternative or complement to traditional expert assessments that could fast-track access to sophisticated sustainability analysis. Perhaps as interesting is the increase in efficiency that our automated process brings. The reduction in assessment time represents a paradigm shift in the scope and frequency with which sustainability assessments can be conducted. This efficiency gain opens up new opportunities for comprehensive, real-time monitoring of urban sustainability initiatives and enables more responsive and adaptive urban planning strategies.

Furthermore, the adaptability of our approach in different contexts - from open citizen data in the Paris participatory budget to specific project information in the context of PROBONO activities - demonstrates its versatility and potential for wide-ranging applications in urban planning and governance. This flexibility suggests that the method could be useful for cities of different sizes, cultural contexts and stages of development. It can offer for example a consistent way to present the alignment of different certification schemes mentioned earlier to the ambitions of the community, allowing the decision makers to opt for the certification scheme that best aligns with their priorities.

Finally, the educational potential of our work should not be underestimated. With reviewers and participants stating that they had a better understanding of the ISO 37101 framework, our research points to innovative ways to bridge the gap between technical sustainability assessments and public engagement.

\section*{Acknowledgement}

This project has received funding from the European Union’s Horizon 2020 Europe Research and Innovation programme under Grant Agreement No 101037075. This output reflects only the author’s view, and the European Union cannot be held responsible for any use that may be made of the information contained therein.

\bibliographystyle{unsrt}  
\bibliography{iso37101_assistant}

\end{document}